\documentclass[%
preprint,
amsmath,amssymb,
aps,
prd,
]{revtex4-2}

\usepackage[colorlinks,bookmarksopen,bookmarksnumbered,citecolor=red,urlcolor=red]{hyperref}

\usepackage{graphicx}
\usepackage{natbib}
\usepackage{color}
\usepackage{latexsym}
\usepackage{subfig}
\usepackage{natbib}
\usepackage{upgreek}
\usepackage{mathrsfs}
\usepackage{float}
\usepackage{caption}
\usepackage{soul}
\usepackage{textcomp}
\usepackage{stmaryrd}
\usepackage{amsmath}
\usepackage{lipsum} 
\usepackage{appendix}
\usepackage{xargs}   
\usepackage[normalem]{ulem}
\usepackage[pdftex,dvipsnames]{xcolor}  
\usepackage[colorinlistoftodos,prependcaption,textsize=small]{todonotes}

\def\der#1#2{{\partial #1\over \partial #2}}

\def\DDt#1{{ \frac{D#1}{Dt}}}

\newcommandx{\eyal}[2][1=]{\todo[linecolor=red,backgroundcolor=red!25,bordercolor=red,#1]{#2}}

\newcommandx{\anirban}[2][1=]{\todo[linecolor=blue,backgroundcolor=blue!25,bordercolor=blue,#1]{#2}}

\def\ol#1{\overline{#1}}
\def\be{\begin{equation}}
\def\ee{\end{equation}}
\def\e{{\rm e}}
\def\ii{{\rm i}}

\usepackage{cleveref}

\crefformat{section}{Sec. #2#1#3} 
\crefformat{subsection}{Sec. #2#1#3}
\crefformat{subsubsection}{Sec. #2#1#3}

\def\be{\begin{equation}}
\def\ee{\end{equation}}

\begin{document}

\preprint{APS/123-QED}

\title{{Pairs of surface wave packets with zero-sum energy in the Hawking radiation analog}}

\author{Anirban Guha}%
 \email{anirbanguha.ubc@gmail.com}

\affiliation{School of Science and Engineering, University of Dundee, Dundee DD1 4HN, UK.}

\author{Eyal Heifetz}
\affiliation{Department of Geophysics, Porter school of the Environment and Earth Sciences, Tel Aviv University,
Tel Aviv 69978, Israel.}

\author{Akanksha Gupta}
\affiliation{Department of Mechanical Engineering,
Indian Institute of Technology, Kanpur, U.P. 208016, India.}




\date{\today}

\begin{abstract} 

Here we propose a minimal analog gravity setup and suggest how to select two surface gravity wave packets in order to mimic some key aspects of Hawking radiation from the horizon of non-rotating black holes. Our proposed setup, unlike the scattering problem conventionally studied, constitutes of a constant mean flow over a flat bathymetry, in which  the two wave packets possess the same amount of wave action but equal and opposite (sign) amount of energy, thereby mimicking virtual particles created out of near horizon vacuum fluctuations. 
Attention is given to the physical mechanism relating  to the signs of the wave action and  energy norm with the wave's intrinsic and total phase speeds. We construct narrow wave packets of equal wave action, the one with positive energy and group speed propagates against the mean flow and escapes from the black hole as Hawking radiation, while the other with negative energy and group speed is drifted by the mean flow and falls into it. Hawking's prediction of low frequency mode amplification is satisfied in our minimal model by construction. We find that the centroid wavenumbers and surface elevation amplitudes of the wave packets are related by simple analytical expressions.    


\end{abstract}

\maketitle


\section{Introduction}
\label{sec:1}

{In the search for laboratory analogs of black-hole radiation (c.f. \citet{Barcelo2019} for an updated review) Schutzhold and Unruh \cite{schutzhold2002gravity}
 theoretically demonstrated how surface gravity waves, in the presence of a counter--current flow in a shallow basin, can be used to simulate phenomena around black holes (BH) in the laboratory.
\citet{rousseaux2008observation} reported  the first successful analog gravity experiment  mimicking white hole (WH) horizons by surface gravity waves. 
\citet{weinfurtner2011measurement} used localized obstacle to block the upstream propagation of a long wave, converting it into a pair of short waves with opposite-signed energy,
one with  positive and the other with negative energy. 
This experiment successfully demonstrated the thermal nature of the stimulated Hawking process at an analog WH horizon. Hawking radiation in analog wave-current systems have been further established experimentally and numerically in recent years, see 
\cite{euve2016observation,robertson2016scattering,euve2020scattering}.
Specifically, \citet{euve2016observation} established analog quantum Hawking radiation
using correlation of the randomly fluctuating free surface downstream of the obstacle.
}

\begin{figure}
    \centering
    \includegraphics[width=0.8\textwidth]{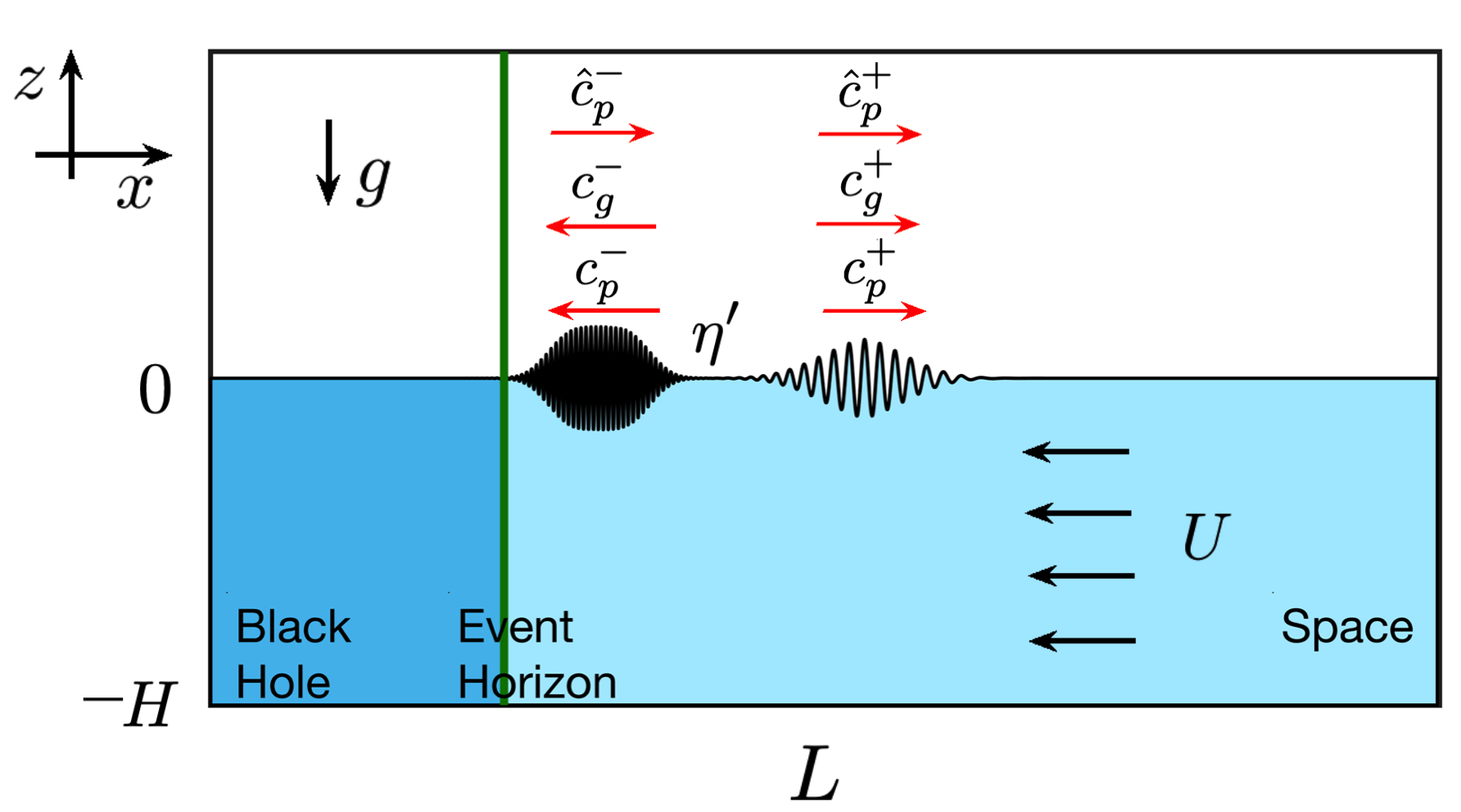}
    \caption{Schematic diagram of the black-hole analog set-up. For details about the various symbols, see text.}
    \label{fig:1}
\end{figure}

{
The objective in this paper is more modest. It aims to propose a minimal water wave analog of pairs of virtual particles with equal and opposite energy, created out of near horizon vacuum fluctuations, where the particle with the positive energy escapes to infinity, and the one with negative energy falls into the BH, leading to BH evaporation \cite{hawking1974black,hawking1975particle}.
As this phenomena by itself is not necessarily related to wave-scattering, it is enough to assume here a flow system with a constant mean counter--current over a flat bathymetry (i.e., constant water depth, see Fig.\,\ref{fig:1}).}

\section{Pseudo-energy and pseudo-momentum}
\label{sec:2}

{Consider for simplicity a rectangular quasi-2D domain $(x,z)$ of the size $(0,L)\times (-H,\eta')$, filled with water (assumed here to be inviscid and incompressible), where $L$ is the horizontal length, $H$ is the mean fluid depth, and $\eta'(x,t)$ denotes the free surface elevation about the mean depth (e.g. Fig.\,~\ref{fig:1}). For this setup the continuity and Euler's momentum equations read: 
\renewcommand{\theequation}{\arabic{equation}a,b}
\be
\nabla\cdot{\bf u} = 0; \qquad \DDt{\bf u} \equiv \left ( \der{}{t} + {\bf u}\cdot\nabla  \right ){\bf u} =  -{\nabla p\over \rho} +{\bf g}\, .
\label{eq:NSE}
\ee
 Here $\nabla \equiv (\partial/\partial x, \partial/\partial z)$ is the 2D gradient operator, ${\bf u} = (u,w)$ denotes velocity, $p$ denotes pressure, $\rho$ is the density of water (assumed constant), and  ${\bf g} = -g{\hat {\bf z}}$ is the gravity vector pointing downwards.}
 
 \renewcommand{\theequation}{\arabic{equation}}
{
Assuming periodic boundary conditions at $x=0$ and $L$,
it is  straightforward to show that both the domain-integrated momentum in the $x$--direction ($P$) and the total fluid energy ($E$): 
\begin{subequations} 
\begin{align}
P   &  = \rho \int_{x=0}^L \int_{z=-H}^{\eta'} u dx dz\,,
\label{eq:PM1}\\
E & =  {\rho\over 2}\int_{x=0}^L \left[ \left ( \int_{z=-H}^{\eta'} |{\bf u}|^2 dz \right) + g\left (\eta'^2-H^2 \right) \right] dx \,,
\label{eq:PE1}
\end{align}
\end{subequations}
are conserved \cite{Buhler2009}. 
The two terms in the RHS of Eq.\,~\eqref{eq:PE1} are respectively the fluid kinetic and potential energy.
Consider a steady mean current in the negative $x$ direction: ${\bf u} = (-\ol{U},0)$ with $\ol{U}>0$, and a constant mean height $H$ satisfying  hydrostatic balance.
This flow is a solution of Eq.\,~\eqref{eq:NSE} and posses the domain integrated momentum and energy 
\renewcommand{\theequation}{\arabic{equation}a,b}
\be
\ol P  = -\rho L H \ol{U}\,,\qquad
\ol E   = {\rho L H \over 2} \left ( \ol{U}^2 - gH \right ) \,.
\ee
\renewcommand{\theequation}{\arabic{equation}}
}
{
Now suppose that on top of this steady base state we add a perturbation that is composed of surface gravity waves of the form 
$\eta'(x,t)=a\e^{\ii(kx-\omega t)}$+ c.c., where $a$ and $k$ respectively denote amplitude and wavenumber (defined positive here), $\omega = k c_p$ denotes frequency, $c_p$ is the phase speed, and c.c. denotes complex conjugate. Then
\be
\omega =   {\hat \omega} -k\ol{U} = k({\hat c}_p - \ol{U}) = k\, c_p \,, 
\label{eq:om1}
\ee
where the intrinsic surface gravity wave frequency and phase speeds (denoted by hat) are given by the familiar dispersion relation:
\be
{\hat \omega} = k{\hat c}_p  = \pm \sqrt{gk\tanh{kH}}\, .
\label{eq:disp_rel}
\ee
Denoting the wave fields by prime so that ${\bf u} = (-\ol{U}+u',w')$, we obtain 
\begin{subequations} 
\begin{align}
P & = \ol P +\delta P\, , \hspace{0.5cm} 
\delta P = \rho \int_{x=0}^L \int_{z=0}^{\eta'} u' dx dz\,, \label{eq:P1} \\
E & = \ol E +\delta E\, , \hspace{0.5cm} 
\delta E =  E' -\ol{U} \delta P\, , \hspace{0.5cm}
E' = {\rho\over 2} \int_{x=0}^L \left ( \int_{z=-H}^{\eta'} {|{\bf u}'|}^2 dz + g{\eta'}^2 \right )dx  \,. \label{eq:E1}
\end{align}
\end{subequations}
}
{
The quantities $\delta P$ and $\delta E$ are  respectively known by (the somewhat confusing terms) pseudo-momentum and pseudo-energy. 
As is evident from Eqs.\,\eqref{eq:P1}--\eqref{eq:E1}, they are simply the momentum and energy contribution of the waves to the system. Since $\ol P$ and $\ol E$ are constant,  $\delta P$ and  $\delta E$ are also conserved (in Appendix \ref{sec:App} we  explicitly show that $\delta E$ in the shallow water limit is equivalent to the energy density integral in \citet[Eqs.~(67--68)]{schutzhold2002gravity}). 
Note that $E'$ -- the positive definite wave eddy energy -- is only one of the contributions by the surface waves to the total change in the energy (as will be clarified further in the next section). Hence, neither the pseudo-momentum nor the pseudo-energy are sign definite; negative pseudo-energy implies that the addition of linear waves to the base flow reduces the energy of the system below its mean value $\ol E$, whereas positive pseudo-energy increases the energy above its mean value. 
}


\section{Pairs of zero-sum Pseudo-energy wave packets}
\label{sec:3}

{
The essential idea in this analogy is that confined surface gravity wave packets represent virtual particles. Therefore we aim to choose superposition pairs of wave packets with equal and opposite values of pseudo-energy $\delta E$ in a way that the sign of their group velocity (in the frame of rest) will be equal to the sign of their pseudo-energy. When this is achieved, the wave packet with the positive pseudo-energy manages to overcome the leftward counter--current $-{\ol U}$ and escapes rightward (from the BH horizon into the outer space), whereas the negative pseudo-energy wave packet is drifted leftward with the base flow (into the BH). 
Consequently, the energy in the left region (inside the BH) is reduced on average and become ${\ol E} -|\delta E|$. Eventually when the leftward wave packet dissipates, it is expected to reduce the  mean energy of BH, so that the new mean energy ${\ol E}_{new} \approx {\ol E} -|\delta E|$.}

\begin{figure}
    \centering
    \includegraphics[width=\textwidth]{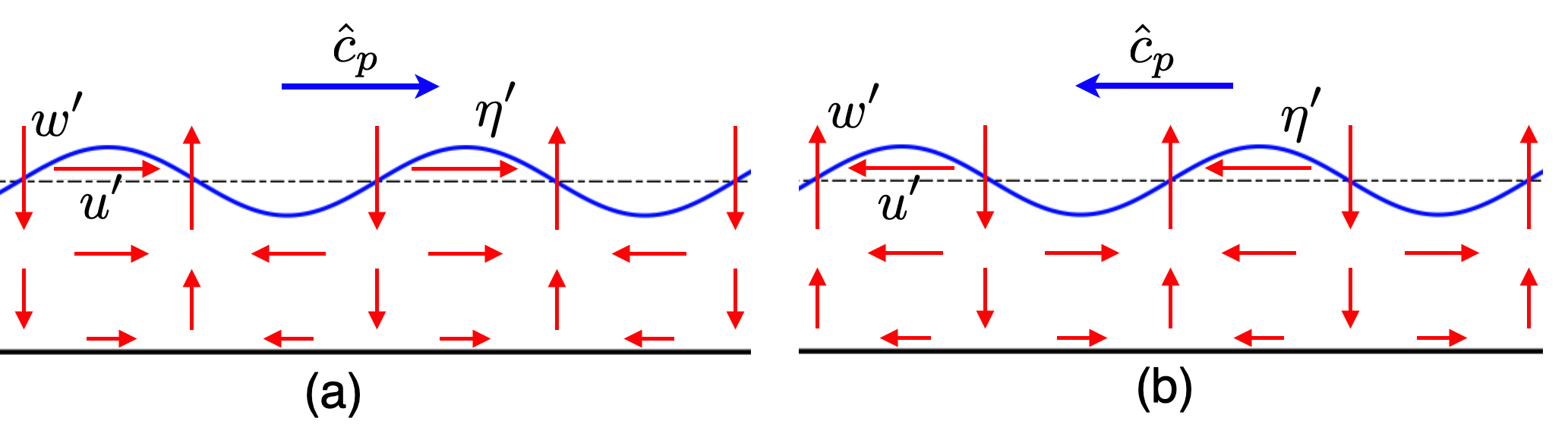}
    \caption{Schematic description of the fact that (a) rightward propagating surface waves have a positive pseudo-momentum, while (b) leftward propagating surface waves have a negative pseudo-momentum.}
    \label{fig:2}
\end{figure}

{Next we wish to suggest how to choose excited pairs of oppositely signed pseudo-energy wave packets based on their physical properties. We first note that for surface  waves it can be shown, after some algebra,  
that the wave eddy energy satisfies:
\be
E'  = {1 \over 2} \rho g L a^2 = {\hat c}_p\,  \delta P\,,
\label{eq:E'}
\ee
implying that ${\hat c}_p$ and $\delta P$ are of the same sign. This sign agreement can be understood from Fig. \ref{fig:2}. The mechanism of surface wave propagation is such that the horizontal convergence (divergence) results in upward (downward) motion that translates the vertical height anomaly $\eta'$. Hence for rightward or positive propagation, ${\hat c}_p > 0$ (Fig. \ref{fig:2}(a)), and $u'$ is in phase with $\eta'$. Therefore the vertical integration of positive $u'$ from the bottom to the wave crests exceeds the vertical integration of negative $u'$ from the bottom to the wave troughs and consequently $\delta P$ is positive, in agreement with Eq.\,\eqref{eq:P1}. By the same argument it follows that $\delta P$ is negative when ${\hat c}_p$ is negative (Fig. \ref{fig:2}(b)).
Equations \eqref{eq:om1}, \eqref{eq:E1},  and \eqref{eq:E'}
then imply  the following relations: 
\be
\delta E = ({\hat c}_p - \ol U)\delta P = c_p \delta P = 
\left (1 - {\ol U \over {\hat c}_p} \right )  E' \, .
\label{eq:PE_def}
\ee} 

Consider then two waves with different wavenumbers $k^+$ and $k^-$ (both defined positive),
where both waves have a positive ${\hat c}_p$ (and hence a positive $\delta P$). Thus both waves are ``trying'' to propagate to the right (in the positive $x$ direction) against the mean current $-\ol U$, see Fig. \ref{fig:1}. If we assume a situation such that
\begin{equation*}
    {\hat c}_p^- < \ol U < {\hat c}_p^+,
\end{equation*}
then  Eq. \eqref{eq:PE_def} implies that $\delta E^+ > 0$ while
$\delta E^- < 0$. In other words, the wave that manages to counter--propagate against the current with a positive phase speed in the rest frame ($c_p^+ > 0$) carries a positive pseudo-energy, whereas the wave whose intrinsic phase speed is not large enough to match the opposed current ($c_p^- < 0$) carries a negative pseudo-energy and consequently propagates to the left in the  rest frame (despite that the  pseudo-momentum of both waves being positive), as shown in Fig. \ref{fig:1}. This statement can be written in terms of frequency and wave-action.
Defining the wave-action as $\delta A \equiv \delta P /k$,
we obtain from Eq. \eqref{eq:PE_def} that $\delta E = \omega \delta A$.
Consider $\delta A$ as an analog for $\hbar$, then for positive $\delta A$ the sign of the pseudo-energy is determined by the sign of its frequency $\omega$.
This suggests that we can set a perturbation of \emph{zero}  pseudo-energy composed of two waves ($\delta E = \delta E^+ + \delta E^- = 0$) with the same positive value of wave-action $\delta A^+ = \delta A^- > 0$. These  in combination yields:
\begin{subequations} 
\begin{align}
\Omega^+ = - \Omega^- > 0 &\implies \hat{\Omega}^++\hat{\Omega}^-= \alpha^+ + \alpha^-,
  \label{eq:omega_nondim}\\
  \bigg(\frac{a^-}{a^+}\bigg)^2 ={ {\hat \Omega}^- \over  {\hat \Omega}^+ } & = 
\sqrt{\alpha^-\tanh{\alpha^-} \over \alpha^+\tanh{\alpha^+}}.
\label{eq:ampratio_nondim}
\end{align}
\end{subequations} 
Here we have used the following non-dimensionalizations: $\alpha^{+(-)} \equiv k^{+(-)}H $, $\hat{\Omega}^{+(-)} \equiv \hat{\omega}^{+(-)} H/\ol U$ 
and $\Omega^{+(-)} \equiv \omega^{+(-)} H/\ol U$. Additionally  Eq. \eqref{eq:om1} has also been used, from which we obtain
$\Omega^{+(-)}=\hat{\Omega}^{+(-)}-\alpha^{+(-)}$, where 
$\hat{\Omega}^{+(-)}=Fr^{-1}\sqrt{\alpha^{+(-)}\tanh{\alpha^{+(-)}}}$, in which the Froude number 
$Fr \equiv \ol U/\sqrt{gH}$.
According to Eq. \eqref{eq:omega_nondim},  the waves have  equal and opposite frequencies. Hence in the rest-frame,   ``$+$'' wave will propagate to the right against the mean current whereas the ``$-$'' wave will be drifted to the left, following the scenario depicted in Fig. \ref{fig:1}. Furthermore, Eq.~\eqref{eq:ampratio_nondim}
provides a direct relation of the amplitude ratio of the ``$+$'' and ``$-$'' waves. 
An interesting point to notice from Eq.~\eqref{eq:ampratio_nondim} is that the condition of zero pseudo-energy superposition does \emph{not} imply that the free surface should be initially flat.

{
While the  pseudo-momentum of a monochromatic sinusoidal wave is perfectly well defined, its position is obviously not. Therefore, in order to generate an initial zero pseudo-energy perturbation whose position and momentum are both reasonably well defined, we should construct pairs of narrow wave packets rather than pairs of monochromatic waves. Hence, the positive (negative) pseudo-energy wave packet should propagate with a positive (negative) group speed $c_g$ (or in non-dimensional terms, ${C_g}^{+(-)} \equiv c_g^{+(-)}/\ol U$), satisfying:
\begin{equation}
    C_g^{+(-)}\equiv \frac{\partial \Omega^{+(-)}}{\partial \alpha^{+(-)}}=-1+\frac{1}{2Fr}\sqrt{\frac{1 }{\alpha^{+(-)}}\tanh \alpha^{+(-)}}\Bigg[1+\frac{2\alpha^{+(-)}}{\sinh 2\alpha^{+(-)}} \Bigg].
    \label{eq:c-g}
\end{equation}
  Furthermore, the centroid group and phase speeds of each wave packet should posses the same sign. This is because the sign of $c_p$ (or in non-dimensional terms, ${C_p}^{+(-)} \equiv c_p^{+(-)}/\ol U$) determines the sign of $\delta E$ whereas the sign of $c_g$ determines the wave packet's direction of propagation.}

\begin{figure}
    \centering
    \includegraphics[width=\textwidth]{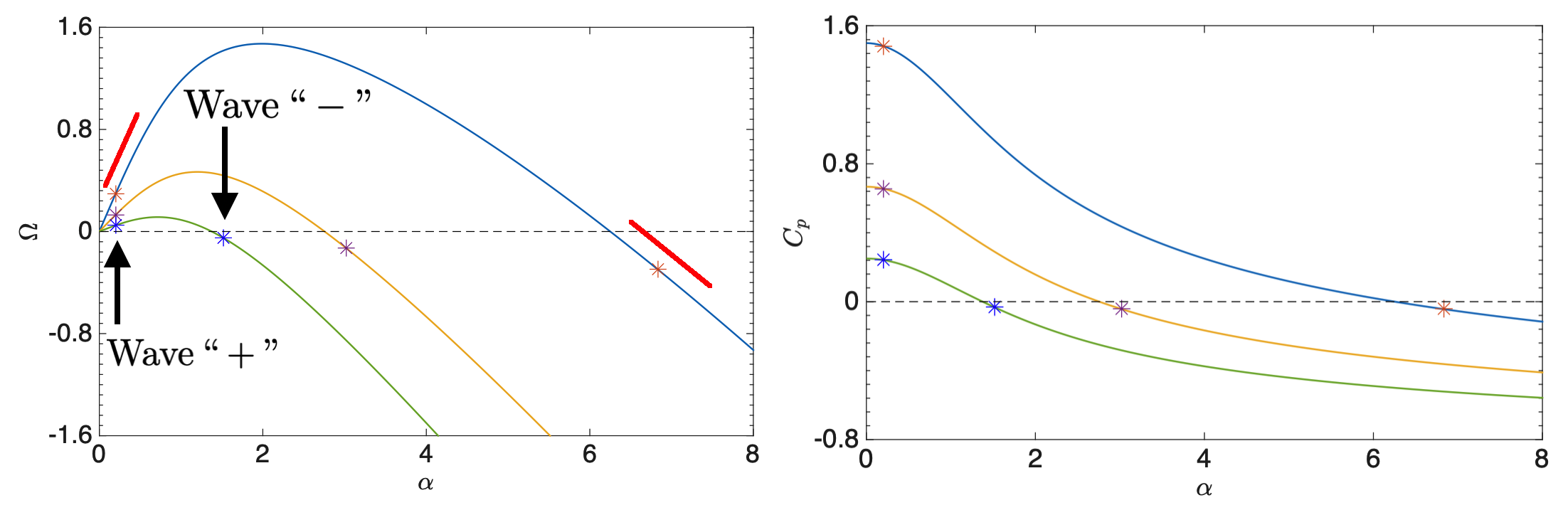}
    \caption{Dispersion curves: (a) $\Omega$ versus $\alpha$, and (b) $C_p$ versus $\alpha$. The blue, yellow and green curves respectively denote $Fr=0.4$, $0.6$ and $0.8$. The short red lines in (a) denotes the slope of the blue curve, which equals to the group speed. The ``\,*\,''s of same color denote a pair-wave; the one above the zero-line has $\delta A>0$ and $\delta E>0$, while that below the  zero-line has $\delta A>0$ and $\delta E<0$.    }
    \label{fig:3}
\end{figure}

{
Consider the positive branch of $\Omega$ and  address only sub-critical flows, i.e. $Fr <1$, in order to enable wave's counter-propagation. The variations of $\Omega$ and $C_p$ with $\alpha$ for different $Fr$ values are respectively plotted in Figs.\, \ref{fig:3}(a) and \ref{fig:3}(b). Two wave packets with equal wave-action, and equal and opposite pseudo-energy, consist of a ``pair-wave'' (denoted by the same colored  ``\,*\,''s), and therefore satisfies Eqs.\, \eqref{eq:omega_nondim}--\eqref{eq:ampratio_nondim}. The ``$+$'' (``$-$'') wave packet's frequency, phase and group speeds are all positive (negative), and hence escapes into space (falls into the BH), in analogy with Hawking radiation. Notice that for sub-critical flows, this condition fails in the shallow-water limit (since the pseudo-energy is always positive); see   Appendix \ref{sec:App}.} 

{Figure \ref{fig:4} shows a pair of wave-packets (both having positive wave-action but equal and opposite pseudo-energy) in a counter-current flow over a flat bathymetry. This configuration is numerically simulated using an in-house High-order Spectral code, detailed in \citet{raj_guha_2019}. As already mentioned, a sum-zero pseudo-energy does not necessarily imply that the superposition of the wave packet pair would render the free surface flat, as clearly shown in Fig.\,\ref{fig:4}(a), which is the configuration at $t=0$. The background flow is sub-critical with $Fr=0.7$. The ``$+$'' wave packet (centroid wavenumber $\alpha^+=0.8$) emits as Hawking radiation while the ``$-$'' wave packet (centroid wavenumber $\alpha^-=2.47$) falls inside the BH; the wave pair has the same magnitude of centroid frequency as per Eq.\,\eqref{eq:omega_nondim}. Here the definition of the event-horizon is arbitrary, however it must be located to the left of the superposed wave packets at $t=0$. The fact that $\alpha^->\alpha^+$ is evident from the dispersion curve in Fig.\,\ref{fig:3}(a). A consequence of $\alpha^->\alpha^+$  is that $a^->a^+$ as per Eq.\,\eqref{eq:ampratio_nondim}, which is also clear from Fig.\,\ref{fig:4}(b). 
}

\begin{figure}
    \centering
    \includegraphics[width=1.0\textwidth]{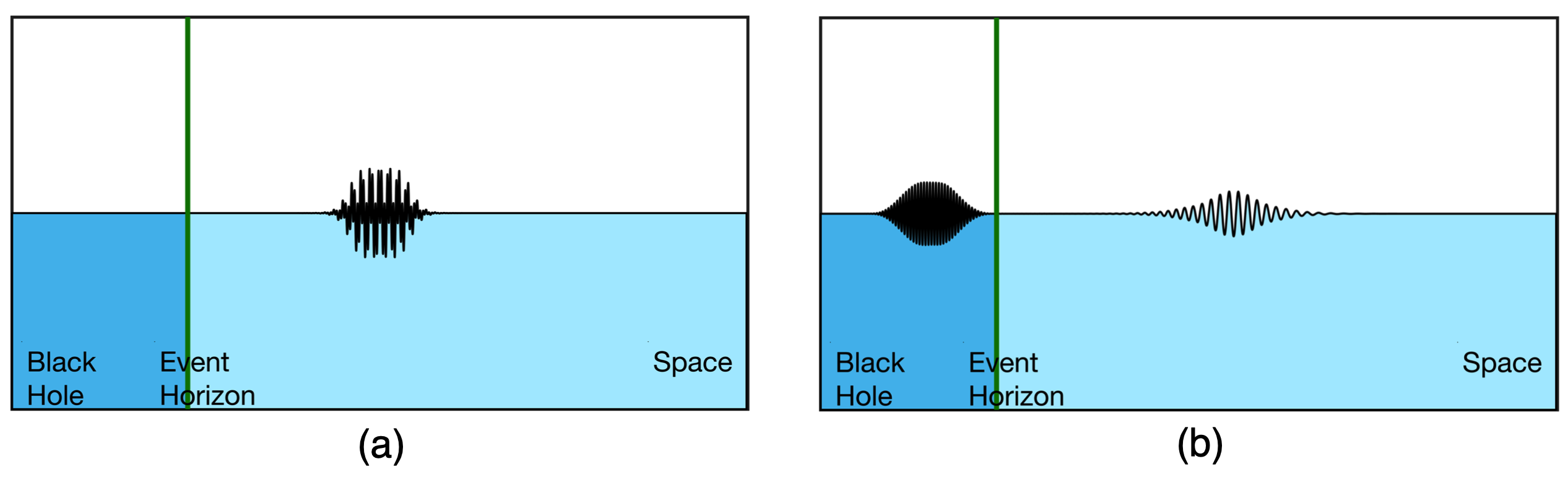}
    \caption{Simulation of sum-zero pseudo-energy wave packet pair for $Fr=0.7$. (a) Configuration at $t=0$, and (b)  configuration at a later time when the ``$+$'' wave packet escapes the BH while the ``$-$'' wave packet falls inside it.}
    \label{fig:4}
\end{figure}

 \section{Parallels with the ratio of Bogoliubov coefficients and low-frequency mode amplification}
{
The study of classical and quantum fields around BHs reveals that a pair wave created with a temporal frequency $\Omega$ satisfies \cite{hawking1974black,schutzhold2002gravity}:
\begin{equation}
\bigg(\frac{\beta^-}{\beta^+}\bigg)^2 =
    \exp{\bigg(-\frac{\Omega}{T}\bigg)},
    \label{eq:Bog}
\end{equation}
where $\beta^{+(-)}$  are referred to as the positive (negative) norm amplitudes (also known as the Bogoliubov coefficients), and $T$ denotes an effective temperature proportional to the surface gravity of a BH. 
According the Hawking's prediction $(\beta^-)^2=
[\exp{(\Omega/T)}-1]^{-1}$, which implies divergence as $\Omega \rightarrow 0$ since for this limit, $(\beta^-)^2 \approx
T/\Omega$.}

{
In analog gravity experiments with surface waves in a counter-current flow over a localized obstacle, parallels between Eq.~\eqref{eq:Bog} and scattering coefficients  were first  established in \citet{weinfurtner2011measurement}, and then in subsequent studies, e.g.\, see Refs.\, \cite{euve2016observation,robertson2016scattering}. Although we have \emph{not} solved a scattering problem here, yet it is interesting to see how the ratio of a conserved norm compares with Eq. \eqref{eq:Bog}. The scattering coefficients in analog-gravity experiments corresponds to the wave-action of the ``$+$'' and ``$-$'' waves \cite{weinfurtner2011measurement}, which in our case are equal by construction (i.e. $\delta A^+=\delta A^-$). Hence the  $\Omega \rightarrow 0$ limit of  Eq. \eqref{eq:Bog} is always satisfied. Furthermore, noting that 
\begin{equation}
  \delta A^{+(-)}=\frac{\rho g L}{2}\,\frac{ \{a^{+(-)}\}^2 }{{\omega}^{+(-)}+k^{+(-)}\overline{U}}\,\,\,,
  \label{eq:wa}
\end{equation}
 we readily find that $\delta A^+ \rightarrow \infty$ when $\hat{\omega}^+ \rightarrow 0$, leading to both $k^+ \rightarrow 0$ and  ${\omega}^+ \rightarrow 0$ (c.f. Fig.\, \ref{fig:3}(a)). Hence by construction $\delta A^- \rightarrow \infty$, however the denominator in Eq.\,\eqref{eq:wa} for this case does not vanish, rather  $a^- \rightarrow \infty$. This fact can also be clearly observed from Eq.\,\eqref{eq:ampratio_nondim}.   }
 
{
In summary, the aspect of low-frequency mode amplification in Hawking's prediction is satisfied by this minimal model.
}

\section{Discussion}
\label{sec:4}

The aim of this paper is to characterize the properties of zero-sum energy pair wave packets in the hydrodynamic analogy of Hawking radiation. First we {wished to clarify} the somewhat non-intuitive physical meaning of positive and negative energy norms (pseudo-energy), how those are related to the wave propagation mechanism, and how the general energy norm converges to the one suggested by \citet{schutzhold2002gravity} in the shallow water limit.  

Next we considered a simple setup consisting of a constant sub-critical counter-current flow over a flat bathymetry; this setup was enough to demonstrate the analog phenomena where positive (negative) energy wave packets escape from (drifted into) the black hole.  The combined requirements of a wave packet pair with equal (and positive in our case)  wave action, and equal and opposite signed pseudo-energy,  determine their centroid wavenumbers as well as their surface elevation amplitude. 

While forming such pairs of wave packets in the laboratory might not be a simple task, it is straight forward to numerically simulate stochastic generation of such zero-sum energy pairs, mimicking near-horizon vacuum fluctuations. The nonlinear effects of wave dissipation and wave-mean flow interaction, which feedback into the counter-current and shift the horizon position, are under ongoing numerical investigation and will be published in a follow-up paper.


\textcolor{orange}{
}

\appendix
\section{Pseudo-energy of shallow water gravity wave} 
\label{sec:App}
\setcounter{equation}{0} \renewcommand{\theequation}{A\arabic{equation}} 

Writing the pseudo-energy explicitly, using Eqs. \eqref{eq:P1}--\eqref{eq:E1} we obtain
\be
\delta E = {\rho\over 2} \int_{x=0}^L \left [ \int_{z=-H}^{\eta'} ({|{\bf u}'|}^2 -2 {\ol U}u')dz + g{\eta'}^2 \right ]dx  \, . 
\label{eq:A1}
\ee
In the shallow water limit, ${|{\bf u}'|}^2 \Rightarrow u'^2$, and $u'$ is not a function of $z$. 
Consequently the pseudo-energy expression for shallow water gravity waves for this set up becomes
\be
\delta E_{SW} = {\rho\over 2} \int_{x=0}^L \left ( H{u'}^2 + g{\eta'}^2 -2 {\ol U}u'\eta' \right )dx  \, . 
\label{eq:A2}
\ee
Let us define the perturbation velocity potential $\phi'$ to satisfy ${\bf u}' = \nabla \phi'$, then for the shallow water the linearized, time--dependent Bernoulli's potential equation (or equivalently, the linearized momentum in the $x$ direction) implies
\be
\left ( \der{}{t} - {\ol U}\,\der{}{x}\right )\phi' =   - g\eta' \, .
\label{eq:A3}
\ee
This relation allows writing the integrand of Eq. \eqref{eq:A2} solely in terms of $\phi'$ 
\be
\delta E_{SW} = {\rho\over 2g} \int_{x=0}^L \left [ g H\left (\der{\phi'}{x}\right )^2 +\left (\der{\phi'}{t}\right )^2  
-\left ({\ol U}\,\der{\phi'}{x}\right )^2 \right ]dx  \, ,
\label{eq:A4}
\ee
which is equivalent to the energy norm defined in Eqs.\, (67)-(68) in \citet{schutzhold2002gravity}. 
Furthermore, for the shallow water surface gravity wave, the amplitudes of the vertical displacement $a$, and the velocity potential amplitude $|\phi|$, are related by \cite{kundu2004}
$$
a = {\alpha|\phi| \over \sqrt{gH}}\,.
$$
Using Eq. \eqref{eq:PE_def} and $\hat{c}_p= \pm \sqrt{gH}$, we can express the pseudo-energy 
in terms of $|\phi|$ as 
\be
\delta E_{SW} = \frac{\rho L}{2 H} \alpha^2 |\phi|^2 (1 \mp Fr).
\label{eq:A5}
\ee
Hence pseudo-energy for shallow-water waves is \emph{always} positive for sub-critical flows ($Fr<1$).
Therefore pairs of opposite pseudo-energy wave packets in sub-critical flows require non shallow water dynamics.

\bibliographystyle{apsrev4-2}
\bibliography{apssamp}

\end{document}